\documentclass[a4paper,12pt]{article}
\usepackage[dvips]{graphicx,psfrag}
\usepackage{amsmath,amssymb}

\def\gev{{\rm GeV}}
\def\mev{{\rm MeV}}
\def\ten{\textbf{10}}

\def\etal{{\it et al.}}

\newcommand{\beq}{\begin{equation}}
\newcommand{\eeq}{\end{equation}}
\newcommand{\bea}{\begin{eqnarray}}
\newcommand{\eea}{\end{eqnarray}}
\newcommand{\bsub}{\begin{subequations}}
\newcommand{\esub}{\end{subequations} \noindent}

\def\PRD#1#2#3{Phys. Rev. {\bf D#1} (#2) #3}

\def\NPB#1#2#3{Nucl. Phys. {\bf B#1} (#2) #3}
\def\PTP#1#2#3{Prog. Theor. Phys. {\bf #1} (#2) #3}

\def\PLB#1#2#3{Phys. Lett. {\bf B#1} (#2) #3}
\def\PRL#1#2#3{Phys. Rev. Lett. {\bf #1} (#2) #3}
\def\PR#1#2#3{Phys. Rep. {\bf #1} (#2) #3}
\begin{document}
\begin{titlepage}
  \begin{flushright}
\begin{tabular}{l}
    {\bf hep-ph/0305016}
\end{tabular}
  \end{flushright}
\begin{center}
    \vspace*{1.2cm}

    {\Large\bf Neutrino Mass Matrix \\
\vspace{0.3cm}
            in Terms of Up-Quark Masses }\\
\vspace{1.0cm}

    {\large
      Masako {\sc Bando}\footnote{E-mail address:
        bando@aichi-u.ac.jp} 
      and Midori {\sc Obara}\footnote{E-mail address:
        midori@hep.phys.ocha.ac.jp}}   \\
    \vspace{4mm}
    $^1$ {\it Aichi University, Aichi 470-0296, Japan} \\[1mm]
    $^2$ {\it Graduate School of Humanities and Sciences, \\ 
Ochanomizu University, Tokyo 112-8610, Japan}
\end{center}
\vspace{1.5cm}
\begin{abstract}
\noindent
We demonstrate that under the  "symmetric zero texture"  with minimal 
Majorana mass matrix, neutrino masses and mixing angles
 are expressed in terms of up-quark masses, 
$m_t,m_c,m_u$. 
This provides interesting relations among neutrino mixing angles 
and up-type quark masses.
Especially  we predict $|U_{e3}| \leq 0.11$ 
even if we include the small mixing effects coming from 
charged lepton side. 
Also absolute masses of three neutrinos are predicted almost uniquely. 
This is quite in contrast to the case where bi-large mixings come from 
the charged lepton sector with non-symmetric charged lepton mass matrix.
\end{abstract}
\end{titlepage}
\section{Introduction}
Recent results from KamLAND~\cite{KamLAND} have 
established the Large Mixing Angle (LMA) solution~\cite{hsmirnov}. 
Combined with the observations  by 
Super-Kamiokande~\cite{kamioka:2002pe,kamioka:2001nj} and 
SNO~\cite{SNO}, this confirms that $V_{MNS}$ has two large mixing 
angles~\cite{Valle,Fogli,Bahcall}:
\begin{equation}
\sin^2 2\theta_{23}  >0.83 ~(99 \%~C.L.), \qquad  
\tan^2 \theta_{12}   = 0.86 \le \sin^2 2\theta_{12}\le 1,
\label{expangle}
\end{equation}
with the mass squared differences 
\begin{equation}
\Delta m^2_{32}   \sim 2.5 \times 10^{-3}~\rm{eV}^2,  \qquad
\Delta m^2_{21}  \sim 7 \times 10^{-5}~\rm{eV}^2. 
\end{equation}
Now the question is  why such a large difference can exist  
 between the quark and lepton sectors. 
Within grand unified theories (GUTs), 
the Yukawa couplings of quarks and leptons to Higgs field are 
related each other. 
Can GUT predict two large mixing angles from some symmetry principle?

The neutrino mixing angles are 
expressed in terms of  MNS matrix~\cite{MNS};   
\begin{equation}
V_{MNS}=U_l^{\dagger}U_{\nu},
\label{MNS}
\end{equation}
where $U_l$ and $U_{\nu}$ diagonalizes $M_l$ and $M_{\nu}$, respectively,  
\begin{eqnarray}
U_l^TM_lU_l = 
{\rm diag}(m_e,m_{\mu},m_{\tau}),  
\quad
U_{\nu}^TM_{\nu}U_{\nu} 
=
{\rm diag}(m_{\nu_e},m_{\nu_{\mu}},m_{\nu_{\tau}}), 
\ \ 
\label{mixing}
\end{eqnarray}
where $M_{\nu}$ is calculated from neutrino  
right-handed Majorana mass matrix ($M_R$)
and Dirac mass matrix ($M_{\nu_D}$); 
\begin{eqnarray}
M_{\nu}=M_{\nu_D}^T M_R^{-1} M_{\nu_D}. 
\label{seesaw}
\end{eqnarray}
We call "up-road" ("down-road") option when 
such a large mixing angle comes from $M_{\nu}$ ($M_l$) side. 
In GUT framework, then, the up- (down-) quark mass matrix 
will play an important role. 
Here we assume the 
up-road option and 
make a semi-empirical analysis by adopting the so-called 
symmetric four zero texture. 
We shall show how we can reproduce 
bi-large mixing angles. 
\section{Symmetric Texture}
First we make a comment on the so-called 
symmetric texture which 
has been extensively investigated 
by many authors~\cite{symtex}.
In general symmetric texture, 
 $M_l$ is hierarchical mass matrix and 
never gives large mixing angles  since down quark mass matrix, $M_d$ 
is hierarchical. Thus symmetric textures dictate only up-road option.   
Can then hierarchical $M_u$ be consistent with $M_{\nu}$ while 
$M_{\nu_D}$ is hierarchical? 
We shall show first that  
large mixing angles does not arise  from 
$M_{\nu}$ if we restrict ourselves to symmetric texture with 
$U(1)$ family structure. 
As we know well, the most popular mechanism which explains 
hierarchical structure of 
masses may be the so-called Froggatt-Nielsen mechanism~\cite{FN} 
using anomalous $U(1)$ family quantum number. 

Let us consider an example  of 2-family model in which we have  
the same $U(1)$ charges to left-handed up-type 
fermions and the right-handed fermions,  
$x_1,x_2$ ($x_1> x_2$), respectively. 
Then we get the form $M_{\nu}$ from Eq.~(\ref{seesaw}) 
with general forms of $M_{\nu_D}$ and $M_R$,
\begin{eqnarray}
M_{\nu_D} \sim 
\left(
\begin{array}{@{\,}cc@{\,}}
      \lambda^{2x_1}  &   \lambda^{x_1+x_2} \\
      \lambda^{x_1+x_2}  &  \lambda^{2x_2}
\end{array}
\right), \quad 
M^{-1}_R
=
\left(
\begin{array}{@{\,}cc@{\,}}
      a       &  c \\
     c       &b 
\end{array}
\right)
\nonumber\\
\rightarrow 
M_{\nu}
\sim (a\lambda^{2x_1}+2c\lambda^{x_1+x_2}+b\lambda^{2x_2})  
\left(
\begin{array}{@{\,}cc@{\,}}
      \lambda^{2x_1}  &   \lambda^{x_1+x_2} \\
      \lambda^{x_1+x_2}  &  \lambda^{2x_2}
\end{array}
\right). 
\label{2times2hie}
\end{eqnarray}
This indicates that $M_{\nu}$ is always 
proportional to the hierarchical matrix, $M_{\nu_D}$. 
Hence,  unless the dominant terms are canceled accidentally 
by making fine tuning, it is impossible to get large mixing  
angles. 

On the contrary, the above argument is no more 
valid if we choose the texture zero matrix in 
Eq.~(\ref{2times2hie}). 
Actually if we take zero texture, we obtain large mixing angle; 
\begin{eqnarray}
&&
  M_{\nu_D}\sim 
\left(
\begin{array}{@{\,}cc@{\,}}
      \lambda^{2x_1}  &   0 \\
     0                 &  \lambda^{2x_2}
\end{array}
\right),  
  M_R \sim 
\left(
\begin{array}{@{\,}cc@{\,}}
      0       &  1 \\
     1       &0
\end{array}
\right)
 \rightarrow  
M_{\nu}
\sim
    \lambda^{2x_i+2x_j}
\left(
\begin{array}{@{\,}cc@{\,}}
    0 & 1 \\
    1 & 0   \
\end{array}
\right).
\nonumber\\
 \label{2times2}
\end{eqnarray}
The above example shows  that 
$M_{\nu}$ is no more proportional to $M_u$ 
(see more general discussion 
in a separate paper~\cite{obarabando}). 
\section{GUT with Symmetric Texture}
The informations of $M_d$ and $M_u$ are well established and 
popular. A simple example of quark mass matrices is  symmetric
  "zero texture " ~\cite{symtex}.   
Let us take the following forms of $M_u$  and $M_d$ which reproduce 
the observed quark and charged lepton masses 
as well as CKM mixing angles~\cite{Nishiura}. 
Then we get their  relations in  $SO(10)$ GUT; 
\begin{eqnarray}
M_U:\quad 
M_u 
\simeq \left(
\begin{array}{@{\,}ccc@{\,}}
0 & \sqrt{m_u m_c} & 0 \\ 
\sqrt{m_u m_c} & m_c & \sqrt{m_u m_t} \\ 
0 & \sqrt{m_u m_t} & m_t 
\end{array}
\right)  
 \leftrightarrow M_{\nu_D}, 
\qquad  
\label{patisalamd}
\end{eqnarray}
Thus once we fix the representation of Higgs 
field in each matrix element,  
$M_l$ and $M_{\nu_D}$ are uniquely determined 
from $M_d$ and $M_u$, respectively. 
\bea
 M_D:\quad 
  M_d 
\simeq  \left(
\begin{array}{@{\,}ccc@{\,}}
0 & \sqrt{m_d m_s} & 0 \\ 
\sqrt{m_d m_s} & m_s &  \sqrt{m_d m_b} \\ 
0 & \sqrt{m_d m_b} &m_b  
\end{array}
\right)  \leftrightarrow M_l.
\label{Mdu}
\eea
Here let us adopt a simple assumption that 
each elements of $M_U$ and $M_D$ is dominated by 
the contribution  either from ${\bf 10}$ or ${\bf 126}$ of $SO(10)$ 
representation. There are 16 options for the Higgs configuration of 
$M_U$ (see Table 
\ref{textureclass}). 
We show that the following option 
of $M_U$ , together with $M_D$ (Georgi-Jarlskog type~\cite{GJ})
and the most economical form of $M_R$; 
\begin{eqnarray}
&&
M_U =
\left(
\begin{array}{@{\,}ccc@{}}
 0                 &{\bf 126}           & 0   \\
{\bf 126}           &{\bf 10}            &{\bf 10} \\
 0                 &{\bf 10}            & {\bf 126}
\end{array}\right), \ 
M_D =
\left(
\begin{array}{@{\,}ccc@{}}
 0                 &{\bf 10}     & 0   \\
{\bf 10}           &{\bf 126}    &{\bf 10} \\
 0                 &{\bf 10}     & {\bf 10} 
\end{array}\right), \ 
\nonumber\\
&&
M_R =\left(
\begin{array}{@{\,}ccc@{\,}}
 0          &  rM_R             & 0   \\
 rM_R          &  0            &0 \\
 0                 &0       & M_R
\end{array}\right)\, .    
\label{massmat}
\eea
can reproduce all the masses and mixing angles of neutrinos 
consistently with present experiments. 
\section{Option S}
Now each matrix element of $M_{\nu_D}$ is determined by 
multiplying an appropriate Clebsch-Gordan (CG) coefficient, 
$1$ or $-3$, and also $M_{\nu}$ are easily calculated 
from  Eq.~(\ref{seesaw}),
\bea
M_{\nu_D} 
=
m_t\left(
\begin{array}{@{\,}ccc@{\,}}
0 & a & 0 \\ 
a & b & c \\ 
0 & c & d
\end{array}
\right), \quad 
M_{\nu}=
\left(
\begin{array}{@{\,}ccc@{\,}}
 0                 &\frac{ a^2}{r}           & 0   \\
\frac{ a^2}{r}   &2\frac{ ab}{r}+ c^2 & c (\frac{a}{r}+1) \\
 0  & c (\frac{a}{r}+1) & d^2
\end{array}
\right) \frac{m_t^2}{m_R} . 
\label{mnu}
\eea
Hence, because of the hierarchical structure of $M_u$, 
$a\ll b\sim c\ll 1$. 
In order to get large mixing angle $\theta_{23}$, the first term 
of 2-3 element of $M_{\nu}$ in Eq.~(\ref{mnu}) should dominate 
and get  the same order of magnitude  as $d^2$, namely, 
$
r \sim \frac{ac}{d^2} \sim  
\sqrt{\frac{m_u^2m_c}{m_t^3}} \sim10^{-7}
$.
This indicates that the ratio of 
the the right-handed Majorana mass of 3rd generation to 
those of the first and second generations, is very small. 
Such kind of mechanism is well known as
"seesaw enhancement"~\cite{Bando:1997ns,tanimoto,smirnov}. 
This tiny $r$ 
 is  very welcome~\cite{Bando:1997ns}; 
the right-handed Majorana mass of 
the third generation  must become of order of GUT scale while those of 
the first and second generations are of order $10^{8}$ GeV. 
This is quite favorable for the GUT scenario to reproduce 
the bottom-tau mass ratio. 
\begin{table}
\caption{Classification of the up-type quark mass matrices.}
{\footnotesize
\begin{center}
\begin{tabular}{|c|r|r|r|r|}
\hline
{} &{} &{} &{} &{}\\[-2ex]
Type & Texture1 &Texture2 & Texture3&Texture4  \\
\hline
$S$ &
$\left(
\begin{array}{@{\,}ccc@{\,}}
0 & \textbf{126} & 0 \\
\textbf{126} & \textbf{10} & \textbf{10} \\
0 & \textbf{10} & \textbf{126} 
\end{array}
\right)^{\mathstrut}_{\mathstrut}$ &
$\left(
\begin{array}{@{\,}ccc@{\,}}
0 & \textbf{126} & 0 \\
\textbf{126} & \ten & \textbf{10} \\
0 & \textbf{10} & \ten 
\end{array}
\right)^{\mathstrut}_{\mathstrut}$ 
&
&
\\ \hline
$A$
&
$\left(
\begin{array}{@{\,}ccc@{\,}}
0 & \textbf{126} & 0 \\
\textbf{126} & \textbf{126} & \textbf{126} \\
0 & \textbf{126} & \textbf{126} 
\end{array}
\right)^{\mathstrut}_{\mathstrut}$ 
& 
$\left(
\begin{array}{@{\,}ccc@{\,}}
0 & \textbf{126} & 0 \\
\textbf{126} & \textbf{126} & \textbf{126} \\
0 & \textbf{126} & \ten 
\end{array}
\right)^{\mathstrut}_{\mathstrut}$
&
$\left(
\begin{array}{@{\,}ccc@{\,}}
0 & \ten & 0 \\
\ten & \ten & \ten \\
0 & \ten & \textbf{126} 
\end{array}
\right)^{\mathstrut}_{\mathstrut}$
& 
$\left(
\begin{array}{@{\,}ccc@{\,}}
0 & \ten & 0 \\
\ten & \ten & \ten \\
0 & \ten & \ten 
\end{array}
\right)^{\mathstrut}_{\mathstrut}$  
  \\  \hline
$B$
&
$\left(
\begin{array}{@{\,}ccc@{\,}}
0 & \ten & 0 \\
\ten & \textbf{126} & \textbf{126} \\
0 & \textbf{126} & \textbf{126} 
\end{array}
\right)^{\mathstrut}_{\mathstrut}$ 
& 
$\left(
\begin{array}{@{\,}ccc@{\,}}
0 & \ten & 0 \\
\ten & \textbf{126} & \textbf{126} \\
0 & \textbf{126} & \ten 
\end{array}
\right)^{\mathstrut}_{\mathstrut}$  
& &
\\  \hline
$C$
& 
$\left(
\begin{array}{@{\,}ccc@{\,}}
0 & \textbf{126} & 0 \\
\textbf{126} & \textbf{10} & \textbf{126} \\
0 & \textbf{126} & \textbf{126} 
\end{array}
\right)^{\mathstrut}_{\mathstrut}$ 
& 
$\left(
\begin{array}{@{\,}ccc@{\,}}
0 & \textbf{126} & 0 \\
\textbf{126} & \textbf{10} & \textbf{126} \\
0 & \textbf{126} & \ten 
\end{array}
\right)^{\mathstrut}_{\mathstrut}$  
&
$\left(
\begin{array}{@{\,}ccc@{\,}}
0 & \ten & 0 \\
\ten & \ten & \textbf{126} \\
0 & \textbf{126} & \textbf{126} 
\end{array}
\right)^{\mathstrut}_{\mathstrut}$ 
& 
$\left(
\begin{array}{@{\,}ccc@{\,}}
0 & \ten & 0 \\
\ten & \ten & \textbf{126} \\
0 & \textbf{126} & \ten 
\end{array}
\right)^{\mathstrut}_{\mathstrut}$  
  \\  \hline
$F$
& 
$\left(
\begin{array}{@{\,}ccc@{\,}}
0 & \textbf{126} & 0 \\
\textbf{126} & \textbf{126} & \textbf{10} \\
0 & \textbf{10} & \textbf{126} 
\end{array}
\right)^{\mathstrut}_{\mathstrut}$ 
&
$\left(
\begin{array}{@{\,}ccc@{\,}}
0 & \textbf{126} & 0 \\
\textbf{126} & \textbf{126} & \textbf{10} \\
0 & \textbf{10} & \ten 
\end{array}
\right)^{\mathstrut}_{\mathstrut}$  
&
$\left(
\begin{array}{@{\,}ccc@{\,}}
0 & \ten & 0 \\
\ten & \textbf{126} & \ten \\
0 & \ten & \textbf{126} 
\end{array}
\right)^{\mathstrut}_{\mathstrut}$ 
& 
$\left(
\begin{array}{@{\,}ccc@{\,}}
0 & \ten & 0 \\
\ten & \textbf{126} & \ten \\
0 & \ten & \ten 
\end{array}
\right)^{\mathstrut}_{\mathstrut}$  
  \\  \hline
\end{tabular}
\end{center}
\label{textureclass}}
\end{table}%
With such a tiny $r$, $M_{\nu}$ is approximately written as, 
\begin{eqnarray}
M_{\nu} 
= \left(
\begin{array}{@{\,}ccc@{\,}}
 0                 &\frac{ a^2}{r}           & 0   \\
\frac{ a^2}{r}   &\frac{2ab}{r} & \frac{ac}{r} \\
 0  & \frac{ac}{r} & d^2
\end{array}
\right) \frac{m_t^2}{m_R}
\equiv  
\left(
\begin{array}{@{\,}ccc@{\,}}
 0                 &\beta           & 0   \\
\beta   &\alpha     & h \\
 0  & h  & 1
\end{array}
\right) \frac{d^2m_t^2}{m_R} , 
\qquad  
\begin{array}{@{\,}c@{\,}}
h=ac/rd^2,  \\
\alpha=2ab/rd^2, \\
\beta= a^2/rd^2, 
\end{array}
\label{apmnu}
\end{eqnarray}
where $\beta \ll \alpha$ and $h\sim \mathcal{O}(1)$. 

First let us diagonalize the dominant term with respect to 2-3 
submatrix of Eq. (\ref{apmnu}) with the rotation angle  $\theta_{23}$,    
\begin{eqnarray}
\stackrel{\longrightarrow }{\theta_{23}}
\ 
\left(
\begin{array}{@{\,}ccc@{\,}}
 0       &\beta \cos\theta_{23}  &\beta \sin\theta_{23}  \\
\beta \cos\theta_{23}  &\lambda_2     &0 \\
\beta \sin\theta_{23}  & 0  &\lambda_{\nu_3}
\end{array}\right), \quad 
\tan^22\theta_{23}&=& \frac{4h^2}{(1-\alpha)^2},  
\label{rot23} 
\end{eqnarray}
with their eigenvalues, 
\begin{equation}
 \lambda_{\nu_3} = \frac{\alpha+1+\sqrt{(\alpha-1)^2+4h^2}}{2},  
\quad  
\lambda_2 = \frac{\alpha+1-\sqrt{(\alpha-1)^2+4h^2}}{2}.
\label{eigenvaluelam}
\end{equation}
Second step is to rotate with respect to 1 and 2 components of
 Eq. (\ref{rot23});
\begin{eqnarray}
&&
\stackrel{\longrightarrow}{\theta_{12}} 
\quad \left(
\begin{array}{@{\,}ccc@{\,}}
\lambda_{\nu_1} & 0 & \beta\sin\theta_{23}\cos \theta_{12}   \\
0 &\lambda_{\nu_2} & \beta\sin\theta_{23}\sin \theta_{12} \\
\beta\sin\theta_{23}\cos \theta_{12} 
 & \beta\sin\theta_{23}\sin \theta_{12}  & \lambda_{\nu_3} 
\end{array}\right), \quad 
\nonumber\\
&&
\tan^22\theta_{12} = 
\Biggl( \frac{2\beta \cos\theta_{23}}{\lambda_2} \Biggr)^2, 
\end{eqnarray}
with eigenvalues,   
\begin{equation}
\lambda_{\nu_2}
=\frac{\lambda_2+\sqrt{\lambda_2^2+4\beta^2\cos^2\theta_{23}}}{2}, 
\quad
\lambda_{\nu_1} 
=\frac{\lambda_2 -\sqrt{\lambda_2^2+4\beta^2\cos^2\theta_{23}}}{2} .
\label{lambdadasshu}
\end{equation}
Finally the neutrino masses are given as, 
\begin{equation}
m_{\nu_3} \sim \lambda_{\nu_3} \frac{d^2m^2_t}{m_R}, \qquad
m_{\nu_2} \sim \lambda_{\nu_2} \frac{d^2m^2_t}{m_R}, \qquad 
m_{\nu_1} \sim \lambda_{\nu_1} \frac{d^2m^2_t}{m_R}.   
\label{applambdadasshu}
\end{equation}
\section{Numerical Calculations} 
Using the up-quark masses at GUT scale within the
error~\cite{koide-fusaoka},  
\begin{equation}
m_u = 1.04^{+0.19}_{-0.20}~\mev, \quad 
m_c = 302^{+25}_{-27}~\mev, \quad
m_t = 129^{+196}_{-40}~\gev,
\label{koidefusaoka2}
\end{equation}
the parameter range of $\alpha=2hb/c$ and $\beta=ha/c$ are estimated from 
the following values; 
\begin{equation}
 \frac{2b}{c}\rightarrow  \frac{2m_c}{\sqrt{m_um_t}} \sim \, 
        1.0-2.4, \qquad 
\frac{\> a \>}{\> c \>}\rightarrow \sqrt{\frac{m_c}{m_t}} \sim 
\,  0.03-0.05.   
\label{abfrm}
\end{equation}
In order to realize large mixing angle  
$\theta_{23}$, the option in which 
$\alpha$ is close to $1$ is a better choice. 
On the other hand, in order to realize large mixing angle  
$\theta_{12}$, $\lambda_2$ 
must become at least of the same order as $2\beta$, so 
the option in which $\beta$ is relatively large 
would be a better choice. 
Thus the desired candidate for the options of Table 1 would be  
1) {\it The Higgs representations coupled with 2-3 and 2-2 elements 
of $M_U$ must be  same.} 
2) {\it The Higgs representation coupled with 1-2 elements of $M_U$ 
must be as large as possible.} 
The option $S$  may be the best candidates which satisfy
the conditions (i) and (ii). 

Leaving the detailed calculations to our full paper~\cite{obarabando}, 
we here show an example of the figures of our results in  Fig.~\ref{fig1}. 
\begin{figure}
\begin{center}
\psfrag{h}{$h$}
\psfrag{sin}{$\sin 2\theta_{23}$}
\includegraphics[width=5.2cm,clip]{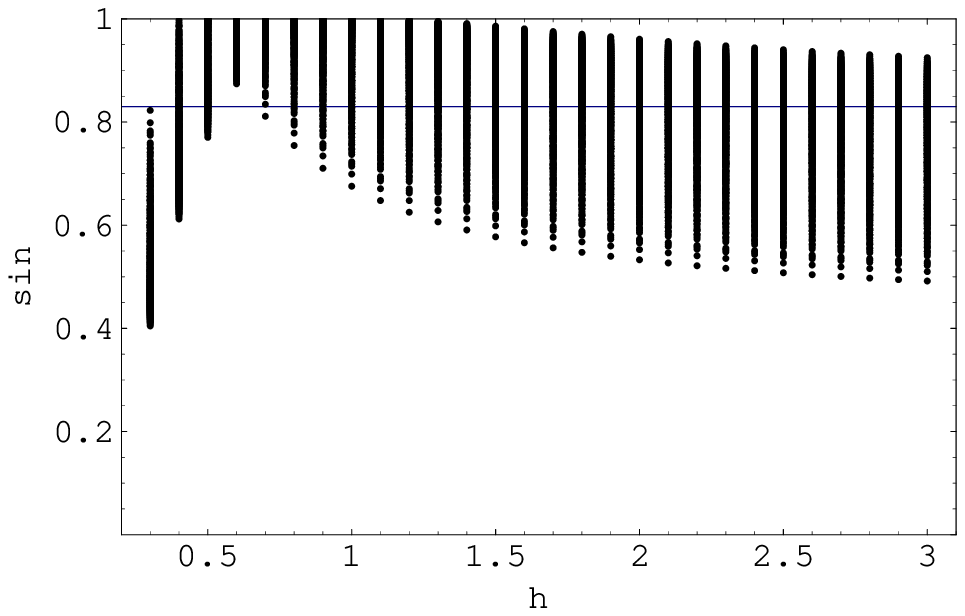}
{\hspace{0.3cm}}
\psfrag{h}{$h$}
\psfrag{tan}{$\tan \theta_{12}$}
\includegraphics[width=5.2cm,clip]{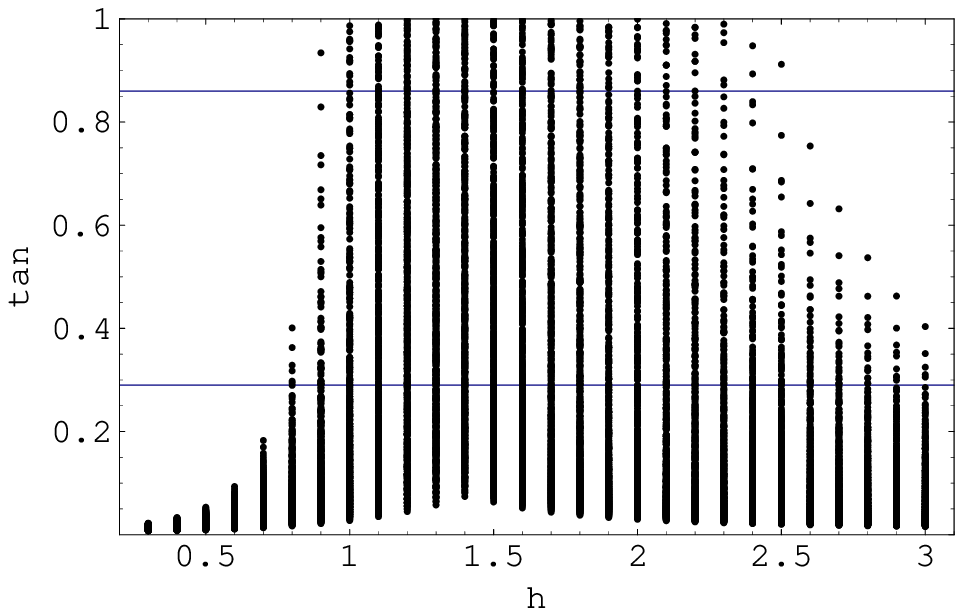}
\end{center}
\vspace{-0.5cm}
\caption{Calculated values of $\sin^2 2\theta_{23}$ 
versus $h$ and $\tan^2 \theta_{12}$ versus $h$ 
in the class $S$.  The experimentally allowed regions are 
indicated by the horizontal lines.} 
\label{fig:Ssin2atm}
\label{fig1}
\end{figure}%
The explicit forms of up-type mass matrix for the class $S$ are 
seen in  Eq.~(\ref{Mdu}) with Eq.~(\ref{massmat}),
which we expected  in section 2. 
Those two types yield the same predictions except for the Majorana mass scale.
The type $S_1$ requires $m_R \sim 2 \times 10^{15}~\gev$ and 
in the type $S_2$, we have $m_R \sim 10^{14}~\gev$, respectively. 
Thus more desirable one may be the type $S_1$ since it 
predicts more realistic bottom-tau ratio at low energy. 
Then the neutrino mass matrix is written as,  
\bea
M_{\nu}
\sim 
\left(
\begin{array}{@{\quad}ccc@{\quad}}
 0  & \ -3h\sqrt{\frac{m_c}{m_t}}\ \  & 0   \\
-3h\sqrt{\frac{m_c}{m_t}} \ & 2h\frac{m_c}{\sqrt{m_um_t}} 
& h \\
 0  &h & 1
\end{array}
\right) \frac{9m_t^2}{m_R}.
\label{app1nuabcr}
\eea
From this, we obtain the following equations, 
\begin{eqnarray}
&&
\tan^22\theta_{23} \simeq  \frac{4h^2}
{ \Bigl(1-h \frac{2m_c}{\sqrt{m_um_t}} \Bigr)^2}\, ,  
\quad
\tan^22\theta_{12} \simeq
\frac{144h^2m_c\cos^2\theta_{23}}
{m_t  \Bigl( 1-2h+h \frac{2m_c}{\sqrt{m_um_t}} \Bigr)^2}. 
\nonumber\\
\end{eqnarray}

The neutrino masses are given by
\bea
m_{\nu_3} \simeq \lambda_{\nu_3} \cdot  \frac{m_t^2}{m_R},\qquad 
m_{\nu_2} \simeq  \lambda_{\nu_2} \cdot  \frac{m_t^2}{m_R},\qquad 
m_{\nu_1} \simeq \lambda_{\nu_1} \cdot \frac{m_t^2}{m_R}, 
\label{neutmasslam}
\eea
where the RGE factor ($\sim 1/3$) has been 
taken account in estimating lepton masses at low energy scale. 
Since $\lambda_{\nu_2} \ll \lambda_{\nu_3} \sim \mathcal{O}(1)$, 
this indeed yields $m_R\sim 10^{15}$ GeV, as many people require. 
We here list a set of typical values of neutrino masses and 
mixings at $h=0.9, \, m_t \simeq 260~\gev$; 
\bea
&\sin^2 2\theta_{23} \sim 0.98-1, \quad \tan^2 \theta_{12} \sim  0.29-0.46, \\
&m_{\nu_3}\sim 0.053-0.059~\rm eV, \quad m_{\nu_2} \sim  0.003-0.008~\rm eV, \\
&m_{\nu_1} \sim 0.0006-0.001~\rm eV. 
\eea
Up to here we have estimated the contribution from $M_{\nu}$ side and 
this is fairly in a good approximation for estimating 
large mixing angles $\theta_{23}$ and $\theta_{12}$. 
Also neutrino masses are determined 
from $M_{\nu}$ only.
However, in estimating $|U_{e3}|$, we  need careful 
estimation, since we cannot neglect the additional contribution 
from the charged lepton side in Eq.~(\ref{MNS}).
The contribution from $M_{\nu}$ is as follows, 
\begin{equation}
\sin\theta_{13} \simeq  \frac{6h}{ 1+2h+h\frac{2m_c}{\sqrt{m_um_t}} }
\sqrt{\frac{m_c}{m_t}}\sin\theta_{23}\cos\theta_{12}.
\end{equation}
Within the allowed range of $h$, the calculated value of 
$\theta_{13}$ from $M_{\nu}$ is almost of the same order as the one from 
$M_l$;
\begin{equation}
|\theta^{\nu}_{13}|  \sim 0.037-0.038 \quad \leftrightarrow  \quad 
\Delta \theta^{l}_{13}\sim \frac{\lambda}{3}\cdot U_{\mu 3}\sim 0.04,
\end{equation}
where $\lambda \sim 0.2$ and the factor 3 comes from 
the Georgi-Jarlskog texture of Eq.~(\ref{massmat}). 
We have to combine those two contributions; 
unfortunately we do not yet have exact information of the relative phase. 
If the two terms act additively (negatively), 
we would have maximal (minimum)  value. 
Still we can say that  $|U_{e3}|$ becomes at most 
$0.11$, which is within the experimental limit~\cite{CHOOZ}.  
This would be one of the very important predictions of this model. 
In order to predict exact $|U_{e3}|$, the inclusion of CP phase of 
$M_{\nu}$ and $M_l$ is important, which is our next task. 
In conclusion we have seen that the up-road option 
can reproduce the present neutrino experimental data very well. 
However also down-road option may be also 
worthwhile to be investigated~\cite{twist}, in which case the Nature 
may show "twisted family structure".  On the contrary 
in the case of up-road option it requires "parallel family structure". 
\section*{Acknowledgements}
We would like to thank to M.~Tanimoto,  A.~Sugamoto and T.~Kugo 
whose stimulating discussion encouraged us very much. 
Also we are stimulated by  the fruitful and instructive discussions 
at the Summer Institute 2002 held at 
Fuji-Yoshida and at the research meeting held in Nov. 2002 
supported by the Grant-in Aid for Scientific Research
No. 09640375. M.~B.\  is supported in part by
the Grant-in-Aid for Scientific Research 
No.~12640295 from Japan Society for the Promotion of Science, and 
Grants-in-Aid for Scientific Purposes (A)  
``Neutrinos" (Y.~Suzuki) No.~12047225, 
from the Ministry of Education, Science, Sports and Culture, Japan.
%
%
%
%

\end{document}